# THE SPIRAL STRUCTURE OF GALAXIES

**Evangelos Chaliasos**

365 Thebes Street

GR-12241 Aegaleo

Athens, Greece

*Abstract*

We start from the author's metric for a rotating and accelerating Universe and we calculate the force acting on a particle in it. This force consists of two parts: one is derived from a potential, and the other describes a Coriolis force coming from the overall rotation of the Universe. We then write down the equations of motion, and we solve them analytically in the Newtonian approximation. The result has to describe in particular a spiral arm. It gives right a logarithmic helix, as it has been found empirically to be the case for our Galaxy (in particular).

**1. Introduction**

In a previous paper [1] the author, motivated by the observation of an amazing similarity between a spiral galaxy and a cyclone on Earth (and since the later is due to the overall rotation of Earth), attempted to find a suitable metric for the Universe, which would describe a rotation of it as a whole. After a couple of special attempts, he finally arrived at the general metric.

To this end he wrote down an appropriate form of the metric describing *two* rotations of the Universe as a whole (assuming it as embedded in a fictitious four-dimensional space), and he derived the resulting Einstein field equations. Then, he solved them analytically, and the result was a new metric of a doubly rotating Universe.

The present paper is dealing with the original motivation underlying ref. [1]. Namely, the author tries to find the equation describing a spiral arm of a galaxy, by finding the motion followed by a particle in the new metric of the Universe, since this is the case for a cyclone on Earth, too. If then this equation agrees with the observations, it will mean that the original assumption of similar causes is correct.

That is, since the spiral structure resembles the motions in a cyclone, then it will be apparent that the similarity will be caused by similar causes, namely the rotation of Earth on the one hand and the rotation of the Universe on the other hand.

Thus, the equations governing the motion of a particle in the new metric of the Universe are derived in the Newtonian approximation in section 2. In section 3 the author finds especially the spatial metric, which then makes possible a suitable renaming of the coordinates, in order to get the right geometrical interpretation. It is then obvious that we are dealing with cylindrical coordinates (in the approximation used). Then the equations are solved analytically in section 4. Finally, in section 5 we derive a single equation describing the spiral arms, which turns out to be a logarithmic helix, which means in agreement with the observations.

**2. The equations**

The metric of the Universe we will use is [1]

$$ds^2 = c^2 dt^2 - U^2 \left[ dz^2 + (d\phi_1 + \omega_1 dt)^2 + z^2 (d\phi_2 + \omega_2 dt)^2 \right], \tag{1}$$

where U, $\omega_1$, $\omega_2$ are taken here to be constants, since they do not have the time to change appreciably compared with the age of the Universe, so that

$$g_{00} = 1 - U^2 \frac{\omega_1^2}{c^2} - U^2 z^2 \frac{\omega_2^2}{c^2} \tag{2}$$

$$g_{01} = 0 \tag{3}$$

$$g_{02} = -U^2 \frac{\omega_1}{2c} \tag{4}$$

$$g_{03} = -U^2 z^2 \frac{\omega_2}{2c}, \tag{5}$$

or, using the notation [4]

$$h \equiv g_{00} \quad \& \quad g_\alpha \equiv -g_{0\alpha}/g_{00}, \tag{6}$$

we may write

$$h = 1 - U^2 \frac{\omega_1^2}{c^2} - U^2 z^2 \frac{\omega_2^2}{c^2} \tag{7}$$

and

$$g_\alpha = \frac{1}{h}\left(0, U^2 \frac{\omega_1}{2c}, U^2 z^2 \frac{\omega_2}{2c}\right). \tag{8}$$

Since we are dealing with a constant field, the force acting on a particle in it is [3]

$$\vec{f} = \frac{mc^2}{\sqrt{1-v^2/c^2}} \left\{ -\nabla \ln \sqrt{h} + \sqrt{h}\frac{\vec{v}}{c} \times \mathrm{curl}\,\vec{g} \right\}. \tag{9}$$

We will work in the Newtonian approximation. Thus we will use the following approximations

$$\vec{f} = \frac{d\vec{p}}{dt} = \frac{d}{dt}\frac{m\vec{v}}{\sqrt{1-v^2/c^2}} \simeq m\ddot{\vec{x}}, \tag{10}$$

where $\mathbf{x} \equiv (z, \varphi_1, \varphi_2)$, and

$$\frac{mc^2}{\sqrt{1-v^2/c^2}} \simeq mc^2. \tag{11}$$

Also

$$h \simeq 1. \tag{12}$$

concerning curl**g**, we have

$$\mathrm{curl}\,\vec{g} = \begin{vmatrix} \hat{i} & \hat{j} & \hat{k} \\ \partial_1 & \partial_2 & \partial_3 \\ g_1 & g_2 & g_3 \end{vmatrix} = -\hat{j}\partial_1 g_3, \tag{13}$$

so that, because of (8), this results in

$$\mathrm{curl}\,\vec{g} = -\hat{j}U^2 z \frac{\omega_2}{c} \tag{14}$$

Now, the first term of **f** (the *potential* term) from (9) gives

$$\vec{f}_1 \simeq -mc^2 \nabla \ln \sqrt{1 - U^2\left(\omega_1^2 + z^2\omega_2^2\right)/c^2}, \tag{15}$$

which results in

$$\vec{f}_1 \simeq U^2 z \omega_2^2 \hat{i}, \tag{16}$$

while the second term (the *Coriolis* force) gives

$$\vec{f}_2 \simeq mc^2 \frac{\dot{\vec{x}}}{c} \times \left(-\hat{j}\partial_1 g_3\right), \tag{17}$$

which results in

$$\vec{f}_2 \simeq U^2 z \omega_2 \hat{j} \times \dot{\vec{x}}. \tag{18}$$

Thus the total acceleration is given by

$$\ddot{\vec{x}} = U^2 z \omega_2^2 \hat{i} + U^2 z \omega_2 \hat{j} \times \dot{\vec{x}}. \tag{19}$$

This is the equations of motion in the Newtonian approximation we were seeking for.

### 3. The spatial metric and renaming of the variables

Writing the metric (1) again, we have [1]

$$ds^2 = c^2 dt^2 - U^2 \left[ dz^2 + (d\phi_1 + \omega_1 dt)^2 + z^2 (d\phi_2 + \omega_2 dt)^2 \right]. \tag{20}$$

The spatial metric can be written as

$$dl^2 = \gamma_{\alpha\beta} dx^\alpha dx^\beta, \tag{21}$$

where the spatial metric tensor $\gamma_{\alpha\beta}$ is defined by

$$\gamma_{\alpha\beta} \equiv -g_{\alpha\beta} + \frac{g_{0\alpha} g_{0\beta}}{g_{00}}. \tag{22}$$

From (20) we have

$$-g_{\alpha\beta} = \begin{vmatrix} U^2 & 0 & 0 \\ 0 & U^2 & 0 \\ 0 & 0 & z^2 U^2 \end{vmatrix}, \tag{23}$$

and we obtain, taking in mind (2) – (5)

$$\gamma_{11} = U^2$$

$$\gamma_{22} = U^2 + \frac{U^4 \frac{\omega_1^2}{4c^2}}{1 - U^2 \frac{\omega_1^2}{c^2} - U^2 z^2 \frac{\omega_2^2}{c^2}}$$

$$\gamma_{33} = z^2 U^2 + \frac{U^4 z^4 \frac{\omega_2^2}{4c^2}}{1 - U^2 \frac{\omega_1^2}{c^2} - U^2 z^2 \frac{\omega_2^2}{c^2}}$$

$$\gamma_{12} = 0$$

$$\gamma_{13} = 0$$

$$\gamma_{23} = \frac{U^4 z^2 \frac{\omega_1 \omega_2}{4c^2}}{1 - U^2 \frac{\omega_1^2}{c^2} - U^2 z^2 \frac{\omega_2^2}{c^2}} \tag{24}$$

Now letting (renaming)

$$\left. \begin{array}{c} z \to r \\ \phi_1 \to z \\ \phi_2 \to \phi \end{array} \right\} \tag{25}$$

and with the values of $\gamma_{\alpha\beta}$ given by the relations (24), the spatial metric (21) can be written as

$$dl^2 = U^2 dz^2 + U^2 dr^2 + r^2 U^2 d\phi^2 + \frac{U^4 r^2 \frac{\omega_1 \omega_2}{2c^2}}{1 - U^2 \frac{\omega_1^2}{c^2} - U^2 r^2 \frac{\omega_2^2}{c^2}} dz d\phi. \qquad (26)$$

Since we work in the Newtonian approximation, we have to take, instead of (26), the metric

$$dl^2 \simeq U^2 \left( dz^2 + dr^2 + r^2 d\phi^2 \right). \qquad (27)$$

We recognize immediately the relation (27) as giving *flat* space in *cylindrical* coordinates.

For the velocity vector we have therefore [5]

$$\dot{\vec{x}} = U(\dot{z}, \dot{r}, r\dot{\phi}), \qquad (28)$$

and for the acceleration vector [5]

$$\ddot{\vec{x}} = U\left( \ddot{z}, \ddot{r} - r\dot{\phi}^2, \frac{1}{r}(r^2 \dot{\phi})^{\cdot} \right). \qquad (29)$$

We have to note that, in (19), **x**′ represents the *contravariant* components of the velocity [4], so that

$$\dot{\vec{x}} = (\dot{z}, \dot{r}, \dot{\phi}), \qquad (30)$$

while in the same equation, (19), **x**″ must be taken as the *covariant* components of the acceleration [4]. Thus, since the common components of the acceleration are given by (29), in order for them to take the covariant ones we have to use the spatial metric (27), so that we have to write

$$\ddot{\vec{x}} = U^2 \left( \ddot{z}, \ddot{r} - r\dot{\phi}^2, (r^2 \dot{\phi})^{\cdot} \right). \qquad (31)$$

### 4. Reduction of the equations

We write (19), if we interchange the positions of the first two coordinates, as

$$U^2 \left( \ddot{z}, \ddot{r} - r\dot{\phi}^2, (r^2 \dot{\phi})^{\cdot} \right) = \left( 0, U^2 r \omega^2, 0 \right) + U^2 r \omega \begin{vmatrix} \hat{i} & \hat{j} & \hat{k} \\ 1 & 0 & 0 \\ \dot{z} & \dot{r} & \dot{\phi} \end{vmatrix}, \qquad (32)$$

from which the following system of simultaneous differential equations results:

$$\ddot{z} = 0 \qquad (33a)$$
$$\ddot{r} - r\dot{\phi}^2 = r\omega^2 - r\omega\dot{\phi} \qquad (33b)$$
$$(r^2\ddot{\phi}) = \omega \dot{r} r \qquad (33c)$$

We see at once that (33a) has the trivial solution

$$z = at + b, \qquad (34)$$

where a and b are constants, and also that z does not enter (33a) & (33c). Thus our system (33) is reduced to

$$\ddot{r} - r\dot{\phi}^2 = r\omega^2 - r\omega\dot{\phi} \qquad (35a)$$
$$(r^2\ddot{\phi})/r = \omega \dot{r} \qquad (35b)$$

If we want to use the usual vector notation, then we have to make use of (28) & (29) for **x′** and **x″** in (19). The vector **g** of course must also be expressed in common components. Then we will take the same system to solve, that is

$$\ddot{r} - r\dot{\phi}^2 = r\omega^2 - r\omega\dot{\phi} \qquad (36a)$$
$$2\dot{r}\dot{\phi} + r\ddot{\phi} = \omega \dot{r} \qquad (36b)$$

### 5. The solution

From (36b) we find

$$r\ddot{\phi} = (\omega - 2\dot{\phi})\dot{r}, \qquad (37)$$

or

$$\frac{d\dot{\phi}}{dt} = (\omega - 2\dot{\phi})\frac{\dot{r}}{r}, \qquad (38)$$

from which

$$\frac{d\dot{\phi}}{\omega - 2\dot{\phi}} = d\ln r, \qquad (39)$$

so that integrading we find finally

$$2\frac{d\phi}{dt} = -\frac{K^2}{r^2} + \omega, \qquad (40)$$

where K is a constant of integration, which has to be taken very small in order to have agreement with the observations.

Then (36a) becomes

$$\frac{\ddot{r}}{r} - \left(-\frac{1}{2}\frac{K^2}{r^2} + \frac{1}{2}\omega\right)^2 = \omega^2 + \omega\left(\frac{1}{2}\frac{K^2}{r^2} - \frac{1}{2}\omega\right), \qquad (41)$$

from which we find

$$\frac{d\dot{r}}{dt} = \left(\frac{1}{4}\frac{K^4}{r^4} + \frac{3}{4}\omega^2\right)r \equiv f(r). \tag{42}$$

Letting r'= p, we then have

$$dp/dt = f(r), \tag{43}$$

or

$$(dp/dr)p = f(r). \tag{44}$$

Integrating, we find

$$p = \pm\sqrt{2\int f(r)dr}, \tag{45}$$

or, performing the integration under the square root, finally

$$p = \pm\sqrt{-\frac{1}{4}\frac{K^4}{r^2} + \frac{3}{4}\omega^2 r^2 + C}, \tag{46}$$

where C is another constant of integration.

Now, since r'= p, we obtain

$$dt = dr/p(r). \tag{47}$$

Substituting p(r) from (46) results in

$$t = \pm\int \frac{dx}{\sqrt{3\omega^2 x^2 + 4Cx - K^4}}, \tag{48}$$

where $x \equiv r^2$. Performing the integration yields

$$t = \pm\frac{1}{\sqrt{3\omega^2}}\ln\left(2\sqrt{3\omega^2 R} + 6\omega^2 x + 4C\right) \tag{49}$$

[3], so that

$$\left\{2\sqrt{3\omega^2(3\omega^2 r^4 + 4Cr^2 - K^4)} + 6\omega^2 r^2 + 4C\right\} = \exp\left(\pm\sqrt{3}\omega t + G\right). \tag{50}$$

Taking all the constants of integration equal to zero (K=C=G=0) we finally find from (50)

$$r = \frac{1}{6\omega}\exp\left(-\frac{\sqrt{3}}{2}\omega t\right). \tag{51}$$

But, we know from (40), that

$$d\phi/dt = (1/2)\omega, \tag{52}$$

so that

$$\phi = \omega t/2. \tag{53}$$

Thus, eliminating t between (51) and (53), we are left with the final result

$$r = (1/6\omega)e^{-\sqrt{3}\phi}, \tag{54}$$

that is we have for the spiral arms the shape of a logarithmic helix, as it is actually observed [2].

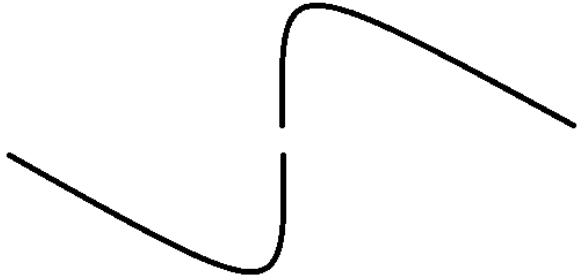